\documentclass[%
 reprint,
 amsmath,amssymb,
 aps,
nofootinbib
]{revtex4-2}

\usepackage{revquantum}
\usepackage{graphicx}
\usepackage{dcolumn}
\usepackage{bm}
\usepackage{upgreek}
\usepackage{xcolor}
\usepackage{enumitem}
\usepackage{placeins}
\usepackage{braket}
\usepackage{siunitx}

\usepackage{dsfont}
\usepackage{comment}

\begin{document}

\title{Optical multi-qubit gate operations on an excitation blockaded atomic quantum register}

\author{Adam Kinos$^1$}
\email{adam.kinos@fysik.lth.se}
\author{Klaus M{\o}lmer$^2$}
\email{moelmer@phys.au.dk}
\affiliation{$^1$Department of Physics, Lund University, P.O. Box 118, SE-22100 Lund, Sweden
}
\affiliation{$^2$Niels Bohr Insitute, Blegdamsvej 17. 2100 Copenhagen, Denmark. 
}
\date{\today}

\begin{abstract}
 We consider a multi-qubit system of atoms or ions with two computational ground states and an interacting excited state in the so-called blockade regime, such that only one qubit can be excited at any one time. Examples of such systems are rare-earth-ion-doped crystals and neutral atoms trapped in tweezer arrays. We present a simple laser excitation protocol which yields a complex phase factor on any desired multi-qubit product state, and which can be used to implement multi-qubit gates such as the n-bit Toffoli gates. The operation is performed using only two pulses, where each pulse simultaneously address all qubits. By the use of complex hyperbolic secant pulses our scheme is robust and permits complete transfers to and from the excited states despite the variability of interaction parameters. A detailed analysis of the multi-qubit gate performance is provided. 
\end{abstract}

\maketitle

\section{\label{sec:intro}Introduction}
Theoretical and experimental efforts have led to immense progress in the implementation of computation on quantum systems. Subject to execution of suitable algorithms, these systems make use of the quantum superposition principle and they may eventually outperform classical computers for many tasks. In a systematic perspective, it has been useful to identify a universal set of one-bit and two-bit gate operations which serve as minimal requirements for the physical implementation of any computational algorithm. But, it has also been recognized that the interaction mechanisms characteristic of each specific physical implementation comes with distinct challenges and advantages. It thus makes sense to carefully choose among formally equivalent but physically different gate operations and sequences of gates that minimize physical resources, execution speed, and errors. This can be done by expert users, and competing automatic and AI inspired strategies are now appearing for such optimization \cite{Mundada2022}.

An especially challenging, while potentially rewarding direction of this research concerns the use of physical interactions between more than two qubits for direct implementation of higher multi-qubit gate operations. This is challenging because it requires analysis of complex physical processes and larger state spaces, and it is, ultimately at variance with the paradigm of breaking computations down to elementary gates. Still, the rewards may be large and, when successful, incorporation of system specific multi-qubit gates in the elementary set, may provide substantial shortcuts and robustness and save computing time. The internal, electronic states that form the qubits in trapped ions all interact simultaneously with the vibrational modes of motion of the ions, and this thus permits implementation of all-to-all effective interactions relevant for quantum simulation \cite{Monroe2021} and multi-qubit conditional gate operations relevant for quantum computing \cite{Martinez2016, Katz2022}. By the Rydberg excitation blockade mechanism neutral atoms interact with all atoms within several micrometre distance and generalization of two-qubit blockade gates \cite{Jaksch2000} can be employed to make multi-qubit Toffoli gates \cite{Isenhower2011} and implement the conditional phase evolution of the Grover algorithm by just few laser pulses \cite{Molmer2011}. Since these specific gates are useful for a wide range of algorithmic tasks and in particular for error correcting codes \cite{Cory1998} and for preparation of pure qubit states \cite{Barber2022}, it is desirable to optimize them and exploit them as much as possible in quantum computing. 

In this article we focus on quantum computing using single rare-earth-ion dopants in inorganic crystals as qubits \cite{Kinos2021}, but our scheme is also applicable to other systems. We combine robust schemes previously explored to enable quantum gates with inhomogeneous ensembles of dopant ions \cite{Roos2004} with the multi-qubit excitation blockade ideas of Ref. \cite{Isenhower2011}, and we assess the expected gate fidelity by numerical simulations and analytical estimates. Compared to implementing single- and two-qubit gate operations in these systems \cite{Kinos2021a}, our protocol only has the additional requirement that all qubits are in the blockade regime and can be addressed simultaneously. In return, our multi-qubit operation can be faster and have smaller errors compared to decomposing a multi-qubit operation into single- and two-qubit operations, while also being more robust against fluctuations in Rabi frequencies and uncertainties in the transition frequencies of the qubits. 

The work is organized as follows. Sec. \ref{sec:sechyp_protocol} presents how our gate operation is performed in a simplified setting and discusses its requirements. The performance of the operation is studied in Sec. \ref{sec:error}. In Sec. \ref{sec:generalization} we generalize the protocol to work with different values of the blockade shifts and to provide phase factors conditioned on any separable multi-qubit state, as well as incorporating single-qubit gates into the execution of the multi-qubit gate. We present a conclusion and outlook in Sec. \ref{sec:conc}.

\begin{figure*}
    \centering
    \includegraphics[width=\textwidth]{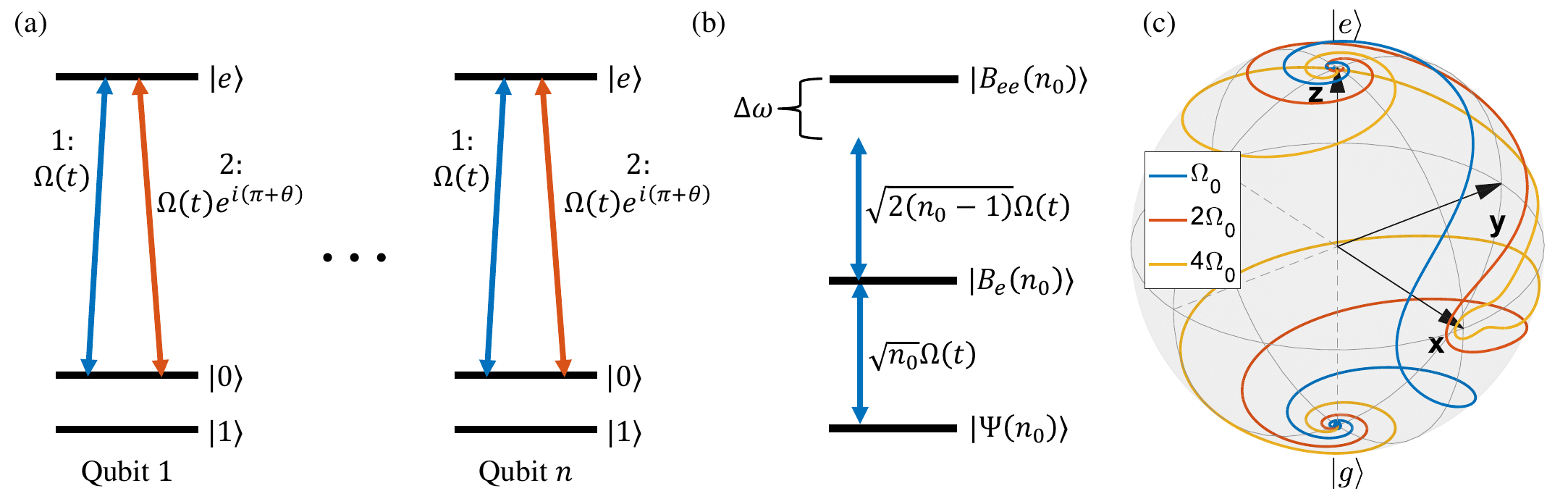}
    \caption{(a) The multi-qubit sechyp operation is applied to $n$ qubit ions that all interact strongly in their excited state $\ket{e}$, e.g., via dipole-dipole or van der Waals interactions. The operation consists of two parts. First, all ions are simultaneously excited by sechyp pulses as described by Eq. (\ref{eq:Sechyp}). Second, the pulses are applied again, except all driving fields have an added phase of $\pi+\theta$ compared to the first pulses. Except for a global phase, this operation applies a phase $\theta$ to the $\ket{11...1}$ state. (b) For a given multi-qubit ground state $\ket{\Psi(n_0)}$ containing $n_0$ $\ket{0}$ components, the Hamiltonian effectively causes excitation, with the interaction strength $\sqrt{n_0}\Omega(t)$, to a superposition state $\ket{B_e(n_0)}$ on the form of Eq. (\ref{eq:Bright}). If all qubits experience the same excited state interaction induced detuning $\Delta\omega$ of its resonance frequency, the state $\ket{B_e(n_0)}$ with a single excitation couples off-resonantly to the state $\ket{B_{ee}(n_0)}$, containing doubly excited state components. For more information see Appendix \ref{app:generalization_AC_err}. (c) Shows trajectories on the Bloch sphere for qubits subjected to sechyp pulses using various Rabi frequencies $\Omega_0$. As can be seen, the sechyp pulse shape can perform complete transfers for different Rabi frequencies, as long as $\Omega_0 \geq \mu\beta$ and $\mu \geq 2$.}
    \label{fig:sechyp_protocol}
\end{figure*}

\section{The multi-qubit sechyp operation}\label{sec:sechyp_protocol}
The goal of our gate operation is to apply a complex phase $\theta$ conditioned on the qubit register populating any separable multi-qubit state. In this section we first present the protocol to apply such a phase $\theta$ on the state $\ket{11...1}$. The case of a general product state and extension to other gates, e.g., the $n$-bit Toffoli gates, is discussed in the subsequent sections.

The system we consider consists of $n$ qubits with two long-lived ground states $\ket{0}$ and $\ket{1}$. As shown in Fig. \ref{fig:sechyp_protocol}(a), we first assume that for each qubit we can choose to apply a laser field that couples the state $\ket{0}$ to the excited state $\ket{e}$. Such individual control can be achieved if the transition frequencies of different qubits are well-separated due to inhomogeneous broadening \cite{Kinos2022a}. Furthermore, we assume that all excited state qubits interact strongly with each other by dipole-dipole or van der Waals interactions, so that if one qubit is excited, it shifts the resonance frequencies of all other qubits and prevents them from being simultaneously excited. This defines the so-called blockade regime.

\begin{figure*}
    \centering
    \includegraphics[width=\textwidth]{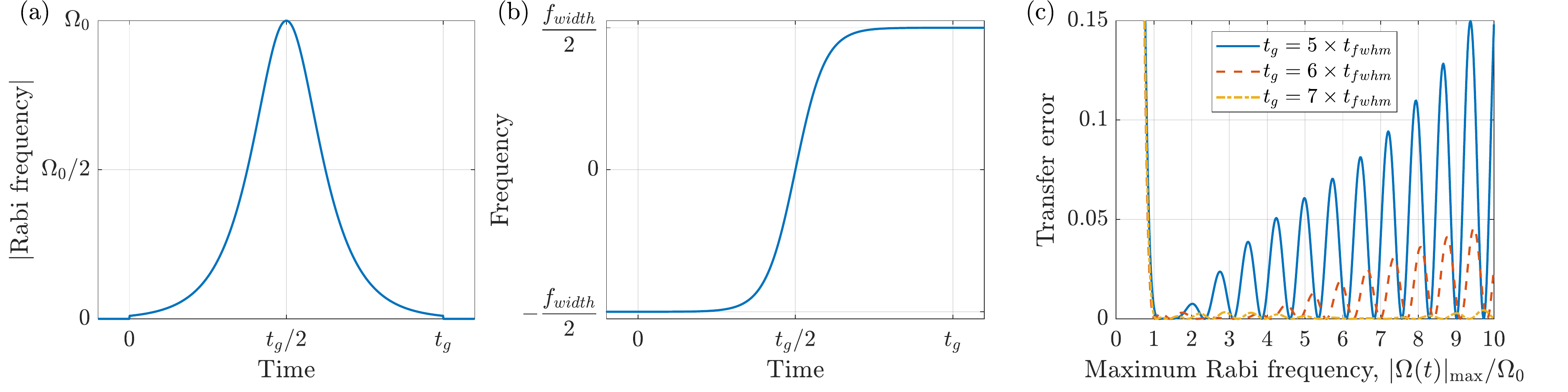}
    \caption{Panels (a) and (b) show the time dependent Rabi frequency amplitude and frequency detuning, respectively, of the sechyp pulse described by Eq. (\ref{eq:Sechyp}). Panel (c) Shows the transfer error as a function of the maximum Rabi frequency amplitude, $|\Omega(t)|_{\text{max}}/\Omega_0$, after performing two consecutive sechyp pulses that, ideally, first excite and then return the ion to its initial state with a $\theta = \pi$ phase shift. For these figures, $\mu = 3$ and $\beta = \Omega_0 / \mu$, which gives a fwhm in intensity of $t_\text{fwhm} = 2\ln{(1+\sqrt{2})} / \beta$ and a frequency width of $f_\text{width} = \mu\beta/\pi$. The cutoff duration is $t_g = 6\times t_\text{fwhm}$ in panels (a-b) and varies in panel (c).}
    \label{fig:sechyp_two_level_error}
\end{figure*}

Our gate operation consists of two applications of the same pulse that simultaneously act on the $\ket{0} \rightarrow \ket{e}$ transition for all $n$ qubits, i.e., the incoming pulse consists of a frequency comb with teeth centered at the transition frequencies of the different qubits that we want to participate in the gate. When qubits are simultaneously addressed like this, a multi-qubit state $\ket{\Psi(n_0)} = \ket{0110...10}$, containing $n_0$ $\ket{0}$ components, couples with a collectively enhanced interaction strength $\sqrt{n_0}\Omega(t)$ to a superposition state
\begin{align}\label{eq:Bright}
    \ket{B_e(n_0)} = \frac{1}{\sqrt{n_0}} \left(\ket{e110...10} + \ket{011e...10} + ... \ket{0110...1e}\right), 
\end{align}
with a single shared excitation among all the qubits that were initially in state $\ket{0}$.
As shown in Fig. \ref{fig:sechyp_protocol}(b), the state $\ket{B_e(n_0)}$ is also off-resonantly coupled with an interaction strength $\sqrt{2(n_0-1)}\Omega(t)$ to $\ket{B_{ee}(n_0)}$, which is a superposition of states with two excited state ions. To make this drive negligible, we assume that the excited state interaction $\Delta\omega$ shifts the resonance enough to suppress excitation of more than a single ion. Thus, for our operation to work $\sqrt{2(n_0-1)}|\Omega(t)|$ and the frequency bandwidth of the sechyp must be much smaller than $\Delta\omega$, for any value of $n_0 = 1,...,n$.

The goal of the operation's first part is to take advantage of the blockade effect to excite all $2^n$ computational multi-qubit ground states except $\ket{11...1}$ to different superposition states that contain exactly one excitation. Thus, the laser pulses must be able to perform complete transfers from $\ket{\Psi(n_0)}$ to $\ket{B_e(n_0)}$ with Rabi frequencies that vary between $\Omega(t)$ for a ground state containing only one $\ket{0}$, to a maximum value of $\sqrt{n}\Omega(t)$, for the state containing $n$ $\ket{0}$. This is accomplished by using a complex hyperbolic secant, or sechyp for short, pulse shape
\begin{align}\label{eq:Sechyp}
    \Omega(t) = \Omega_0 \text{sech}\left(\beta \left(t - \frac{t_g}{2}\right)\right)^{1+i\mu}, 
\end{align}
which is robust against variations of the overall Rabi frequency as long as $\Omega_0 \geq \mu \beta$ and $\mu \geq 2$ \cite{Silver1985}, as indicated in Fig. \ref{fig:sechyp_protocol}(c). An example of the Rabi frequency amplitude and frequency of a sechyp pulse is shown in Fig. \ref{fig:sechyp_two_level_error}(a-b). An added benefit of using sechyp pulses is that they are also robust against variations in transition frequencies \cite{Roos2004}. 

The second part of the operation is identical to the first one, except that all driving fields are applied with a phase changed by the constant amount $\pi+\theta$ compared with the first pulse, as shown in Fig. \ref{fig:sechyp_protocol}(a). This will thus deexcite all state components excited by the first pulse back to their respective ground state with a phase shift of $-\theta$. Except for a global phase, this operation is equivalent to applying a phase of $\theta$ to the qubit register state $\ket{11...1}$. 

\section{Gate performance}\label{sec:error}
In this section we investigate the performance of the gate operation and its robustness against three error sources: the error due to imperfect sechyp transfers; the error due to the off-resonant coupling to the doubly excited states $\ket{B_{ee}(n_0)}$; and the error due to $T_2$ dephasing of the excited state. 

We assume here that all qubits interact with the same interaction shift given by $\Delta\omega$, but we shall return to this issue again in Sec. \ref{sec:arb_shifts}. Furthermore, we assume that the initial state is an even superposition of all $2^n$ computational ground states, and defer discussion of the general case and some analytical results to Appendix \ref{app:error_est}. Under these assumptions, the total error can be estimated as
\begin{widetext}
\begin{align}\label{eq:total_error}
    \epsilon &= 1 - \frac{1}{2^{2n}}\bigg(1 + \sum_{n_0=1}^{n} \binom{n}{n_0}2 e^{-\gamma} \text{Re}\left[A(n_0)\right] + \nonumber\\ 
    &\sum_{n_0=1}^{n}\sum_{m_0=1}^{n} \sum_{k=\text{max}(n_0+m_0-n, 0)}^{\text{min}(n_0, m_0)} \text{Re}\left[A(n_0)A^*(m_0)\right] \binom{n}{n_0} \binom{n_0}{k} \binom{n-n_0}{m_0-k} \frac{k + (n_0 m_0 - k)e^{-2\gamma}}{n_0 m_0} \bigg),
\end{align}
\end{widetext}
where $\gamma = \alpha t_g / T_2$ represents the dephasing error during the pulse duration $t_g$ due to the finite coherence time $T_2$ ($\alpha \approx 1$ estimates how large fraction of the pulse duration the atom spends in the excited state). $A(n_0)$ are complex numbers 
\begin{align}\label{eq:A_n0}
    A(n_0) &= T(n_0)\text{exp}\left(i\frac{2(n_0-1)\Lambda}{4\Delta\omega}\right), \nonumber \\
    \Lambda &= \int_0^{t_g} |\Omega(t)|^2 dt = \frac{2\Omega_0^2}{\beta} \text{tanh}\left(\beta t_g/2\right),
\end{align}
and represent the effect of imperfect state transfer, $T(n_0)$, and an AC Stark shift of the singly excited state due to the off-resonant coupling to higher excited states, which is discussed further in the subsequent subsections. Finally, the pulse and interaction parameters $\Omega_0$, $\Delta\omega$, and $\beta$ are all given in angular frequency units.

\subsection{Transfer errors}\label{sec:transfer}
The transfer of state amplitude to and from the excited states is not perfect, and errors occur because the sechyp pulse has a cutoff duration $t_g$ which leads to small jumps in the Rabi frequency amplitude at $0$ and $t_g$. The transfer error increases if these jumps are larger, which occurs either if the overall Rabi frequency is increased or if the cutoff duration is reduced, as shown in Fig. \ref{fig:sechyp_two_level_error}(c). For the multi-qubit operation, the different ground states $\ket{\Psi(n_0)}$ are driven with different Rabi frequencies $\sqrt{n_0}\Omega(t)$ where $n_0 = 1,...,n$, and the transfer error is thus different for different ground state components. $T(n_0)$ in Eq. (\ref{eq:total_error}) is related to the transfer error in Fig. \ref{fig:sechyp_two_level_error}(c) via $\epsilon_{\text{transfer}} = 1 - |T(n_0)|^2$, and for more information about the calculation of $T(n_0)$ see Appendix \ref{app:simulation}. 

To reduce the transfer error one can increase the pulse duration, $t_g$, before the cutoff of the laser field. Alternatively, one can modify the sechyp pulse shape to smoothly approach zero at the start and end of the pulse. This may secure better convergence to the adiabatic eigenstate of the Hamiltonian at the end of the pulse, but we deem this to be outside the scope of this paper.

\subsection{AC Stark shift errors}\label{sec:AC}
The operation requires that the drive between $\ket{B_e(n_0)}$ and $\ket{B_{ee}(n_0)}$ as shown in Fig. \ref{fig:sechyp_protocol}(b) is negligible. If all qubits interact with the same blockade shift $\Delta\omega$, the transition is driven off-resonantly with a collectively enhanced interaction strength $\sqrt{2(n_0-1)}\Omega(t)$. If this and the bandwidth of the pulse are both much less than $\Delta\omega$, the effect of the drive can be modeled as an AC Stark shift of the $\ket{B_e(n_0)}$ state given by
\begin{align}\label{eq:AC_Stark}
    \omega_{\text{AC}}(t) = \frac{2(n_0-1)|\Omega(t)|^2}{4\Delta\omega}.
\end{align}
This frequency shift introduces a continuous phase error when the system occupies the state $\ket{B_e(n_0)}$, which occurs for roughly the duration, $t_g$, of one of the sechyp pulses. The AC Stark shift error is therefore modeled through the second factor of $A(n_0)$ in Eq. (\ref{eq:A_n0}). 

One can reduce the AC Stark shift error by reducing $\Omega_0$ or the cutoff duration, or by increasing $\Delta\omega$.

\subsection{Dephasing errors}\label{sec:dephasing}
Excited states typically have worse coherence properties compared to the computational ground states, and in this work we investigate errors due to pure dephasing in the excited state. While the error analysis presented in Appendix \ref{app:error_est} is straightforward to generalize to other dephasing models, we here assume that the excited state components dephase independently of each other, such that $\ket{e1}$ dephases with $\ket{11}$ with a time constant $T_2$, $\ket{e0}$ dephases with $\ket{0e}$ with a time constant $T_2/2$, while $\ket{e0}$ does not dephase with respect to $\ket{e1}$. When estimating the total error using Eq. (\ref{eq:total_error}) dephasing is included through $\gamma$. 

For a given $T_2$ one can reduce dephasing errors by reducing the duration of the pulse, i.e., increasing $\Omega_0$ or reducing the cutoff duration.

\begin{figure*}
    \centering
    \includegraphics[width=\textwidth]{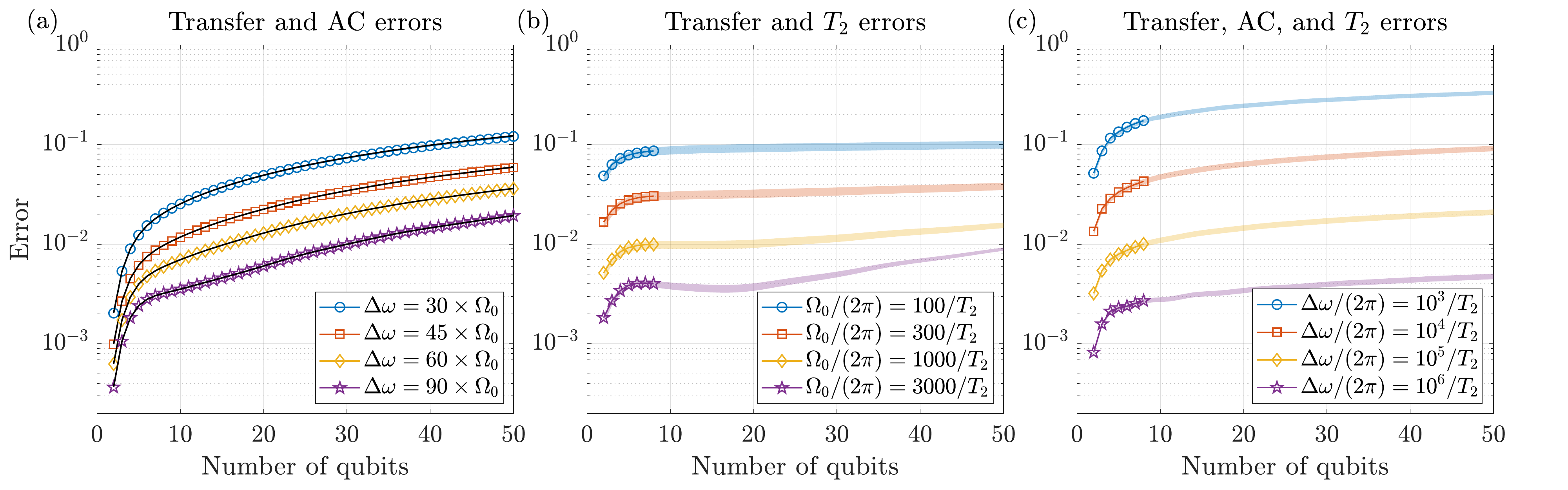}
    \caption{This figure shows the different error contributions. (a) transfer and excited state AC Stark shift error, (b) transfer and dephasing errors, and (c) transfer, AC, and dephasing errors, as a function of the number of qubits in the multi-qubit operation. The panels show results for different values of $\Delta\omega / \Omega_0$ in (a), $\Omega_0 T_2$ in (b), and $\Delta\omega T_2$ in (c). Numerical results obtained by solving the Schr\"odinger equation for $n=2,...,50$ qubits in (a) and the Lindblad master equation for $n=2,...,8$ qubits in (b-c), are shown by markers. For information regarding the simulations, see Appendix \ref{app:simulation}. The estimated errors obtained from Eq. (\ref{eq:total_error}) are shown by (a) black solid lines, (b-c) colored regions where the uncertainty comes from using $\alpha = 0.9 \rightarrow 1.1$ in $\gamma = \alpha t_g / T_2$ to estimate the duration spent in the excited state. The initial state is an even superposition of all computational ground states, and the operation uses $\theta = \pi$. The sechyp parameters are the same as in Fig. \ref{fig:sechyp_two_level_error}, with $t_g = 6 \times t_{\text{fwhm}}$ in (a-b), whereas $t_g / t_{\text{fwhm}}$ in (c) is optimized together with $\Omega_0$ to minimize the error of Eq. (\ref{eq:total_error}). Here we assume that all qubit pairs interact with the same blockade shift $\Delta\omega$, but Fig. \ref{fig:random_shifts_err} shows the more general case.}
    \label{fig:sechyp_errors}
\end{figure*}

\subsection{Results}\label{sec:results}
We first study the multi-qubit gate error when no dephasing is present, i.e., only errors due to imperfect transfers and AC Stark shifts are considered. Figure \ref{fig:sechyp_errors}(a) shows the numerical error obtained by solving the Schr\"odinger equation and the error estimated by Eq. (\ref{eq:total_error}), as a function of the number of qubits $n$ for different ratios of $\Delta\omega/\Omega_0$. The main error stems from the AC Stark shift, but when $\Delta\omega/\Omega_0$ is high the sechyp pulse transfer errors also contribute significantly. As expected, the error grows with increasing $n$ and with decreasing values of $\Delta\omega/\Omega_0$. If $\Delta\omega/\Omega_0$ is doubled the AC Stark shift error is reduced by roughly a factor four, which can be understood from Eqs. (\ref{eq:total_error}) and (\ref{eq:A_n0}) since $\Lambda/\Delta\omega \propto \Omega_0/\Delta\omega$ (as $\beta = \Omega_0 / \mu$) and $\text{Re}[\text{exp}(i\phi)]\approx 1 - \phi^2/2$ when $\phi = \frac{2(n_0-1)\Lambda}{4\Delta\omega}$ is small. Interestingly, the error scales almost linearly with $n$, despite $\phi \propto (n_0-1)$ and $n_0 = 1,...,n$. The reason is that most terms in Eq. (\ref{eq:total_error}) have $n_0 \approx n/2$, and since they all obtain similar phase shifts their relative phases only scale as $n$. If, instead, the initial multi-qubit state is a GHZ state, $\ket{00...0} + \ket{11...1}$, the error grows as $n^2$ as expected. Thus, the AC Stark shift error of the operation depends on the initial state, and this can be estimated using Eq. (\ref{eq:total_error_all_init}) in Appendix \ref{app:error_est}. 

We now investigate the effect of transfer and dephasing errors and assume infinite blockade shifts such that no AC Stark shift errors occur. The error obtained by solving the Lindblad master equation and the estimated error can be seen in Fig. \ref{fig:sechyp_errors}(b). When $n$ or $\Omega_0 T_2$ are small, the error is mainly due to dephasing, and if $\Omega_0 T_2$ is increased the error decreases by roughly the same factor. Furthermore, for $n \sim 10$ the error from dephasing begins to saturate. This happens because the main error in Eq. (\ref{eq:total_error}) for large $n$ stems from the last term and since $n_0 m_0$ grows faster than $k$, most components obtain the same factor $e^{-2\gamma}$, and thus the error due to dephasing tend toward $1 - e^{-2\gamma}$. However, if $n$ and $\Omega_0 T_2$ are large, the errors due to the imperfect transfers become non-negligible, and the corresponding error curves do not saturate. The transfer error oscillates as a function of Rabi frequency as shown in Fig. \ref{fig:sechyp_two_level_error}(c), and this explains the non-monotonic behavior of the error as a function of $n$ for the highest value of $\Omega_0 T_2$ shown in Fig. \ref{fig:sechyp_errors}(b). 

Finally, in Fig. \ref{fig:sechyp_errors}(c) $\Omega_0$ and the cutoff duration $t_g$ are optimized to minimize the total error of Eq. (\ref{eq:total_error}) for different values of the product $\Delta\omega T_2$. If we neglect the transfer error and optimize the cutoff duration, the error due to the AC Stark shift scales as $(\Omega_0/\Delta\omega)^2$ and the error due to dephasing scales as $1/(\Omega_0 T_2)$. The total error should therefore scale roughly as 
\begin{align}
    \epsilon \propto \left(\frac{\Omega_0}{\Delta\omega}\right)^2 + \frac{C(n)}{\Omega_0 T_2},
\end{align} 
where $C(n)$ contains the relative scaling factor between the two terms, which in general depends on the number of qubits $n$. When minimized with respect to $\Omega_0$ this yields 
\begin{align}
    \Omega_0 = \left(\frac{C(n) \Delta\omega^2}{2T_2}\right)^{1/3},
\end{align}
which gives 
\begin{align}
    \epsilon \propto \left(\frac{1}{\Delta\omega T_2}\right)^{2/3}.
\end{align}
Thus, if $\Delta\omega T_2$ increases by a factor of ten, the error is reduced by a factor of roughly $10^{2/3} \approx 4.6$, which is in good agreement with the numerical results in Fig. \ref{fig:sechyp_errors}(c).

\section{Generalizations of the multi-qubit sechyp operation}\label{sec:generalization}
In this section we make three generalizations of our gate operation: we allow the blockade shift to differ between different qubit pairs; we show that the protocol can be used to apply any phase to any separable multi-qubit state; and we show how the operation may readily incorporate single-qubit gates and thus shorten quantum circuits with combined single- and multi-qubit gates.

\subsection{Arbitrary blockade shifts}\label{sec:arb_shifts}
If the blockade shifts, $\Delta\omega_i$, are different for each qubit pair $i$, state $\ket{B_{ee}(n_0)}$ is no longer decoupled from the other doubly excited state components. However, the AC Stark shift error can still be estimated using Eqs. (\ref{eq:total_error}) and (\ref{eq:A_n0}), except that $\Delta\omega$ now depends on the multi-qubit state $\ket{\Psi(n_0)}$ and is replaced by an effective shift $\Delta\omega_{\text{eff}}$ of state $\ket{B_{ee}(n_0)}$, which can be calculated through the recursive Eq. (\ref{eq:recursive}) presented in Appendix \ref{app:generalization_AC_err}. However, the estimation of this error requires the calculation of roughly $2^n$ different $\Delta\omega_{\text{eff}}$, which is not feasible when $n$ is large. 

\begin{figure}
    \centering
    \includegraphics[width=\columnwidth]{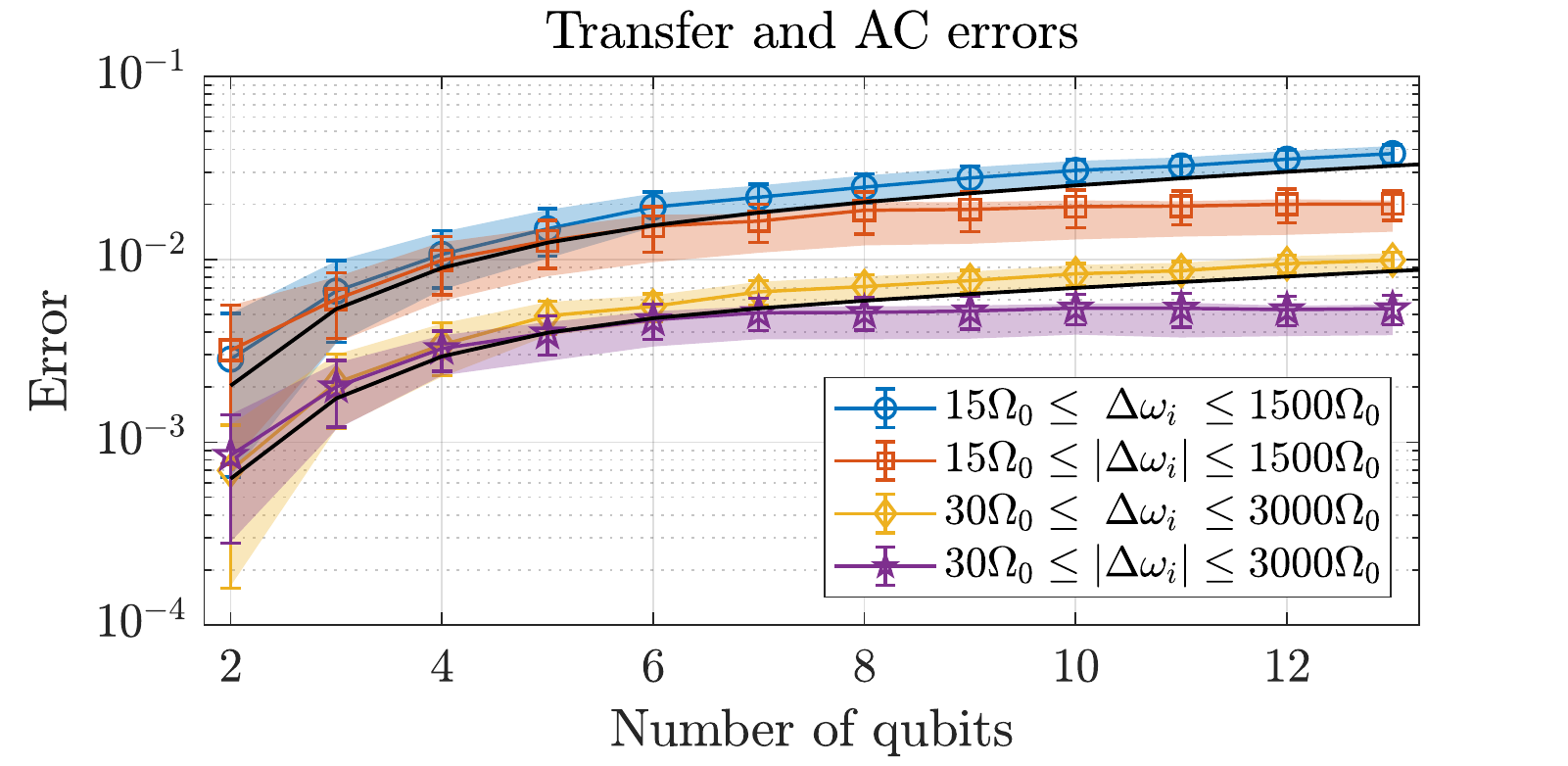}
    \caption{The figure shows the transfer and AC Stark shift errors as a function of the number of qubits when the blockade shifts, $\Delta\omega_i$, are different for each qubit pair $i$. The average error $\pm$ one standard deviation is obtained by sampling 100 randomizations of all blockade shifts, and is shown in colored markers for numerical simulations and colored regions for the theoretical model presented in Appendix \ref{app:generalization_AC_err}. When all $\Delta\omega_i$ are positive, the AC Stark shift error can be estimated using an average shift for all qubits, i.e., the results of Fig. \ref{fig:sechyp_errors}(a) can be used with $\Delta\omega_{\text{avg}} \approx 30 \Omega_0$ (yielding the top solid black line) and $\Delta\omega_{\text{avg}}\approx 60\Omega_0$ (yielding the bottom solid black line), respectively, for the ranges $15\Omega_0 \leq \Delta\omega_i \leq 1500\Omega_0$ and $30\Omega_0 \leq \Delta\omega_i \leq 3000\Omega_0$. When the shifts are both positive and negative, the AC Stark shift error saturates when $n \sim 10$.}
    \label{fig:random_shifts_err}
\end{figure}

Figure \ref{fig:random_shifts_err} shows the results of both numerical simulations and theoretical estimations of the AC Stark shift errors for up to $n = 13$ qubits. The individual blockade shifts lie in the range $\Delta\omega_\text{min} \leq \Delta\omega_i \leq \Delta\omega_\text{max}$, where the inverse of the shifts, $\frac{1}{\Delta\omega_i}$, are randomly drawn from the uniform distribution $\left[\frac{1}{\Delta\omega_\text{max}}, \frac{1}{\Delta\omega_\text{min}}\right]$, to model that the shifts vary as $1/r^3$ where $r$ is the distance between the two qubits and the qubits are distributed evenly in three dimensions. 

If all $\Delta\omega_i$ are positive, the average AC Stark shift error can be estimated using one average effective shift $\Delta\omega_{\text{avg}} = 1/\langle \frac{1}{\Delta\omega_i}\rangle$ for all qubit pairs, i.e., one does not have to use the recursive formula presented in Appendix \ref{app:generalization_AC_err}, and the results of Fig. \ref{fig:sechyp_errors} apply if one replaces $\Delta\omega$ with $\Delta\omega_{\text{avg}}$. 

If $\Delta\omega_i$ take both positive and negative values, while the absolute values lie within the previously listed range, the off-resonant driving to the doubly excited states can induce both positive and negative phases which leads to a saturation in the AC Stark shift error when $n \sim 10$. 

\subsection{Applying any phase to any separable multi-qubit state}\label{sec:arb_gate}
So far the qubits were driven only on their $\ket{0}\rightarrow\ket{e}$ transitions with Rabi frequencies $\Omega^{0e}(t) = \Omega(t)$, which applies a phase $\theta$ to $\ket{11...1}$. We now generalize the protocol to apply a phase $\theta$ to any separable multi-qubit state. For each qubit $q$ we apply fields on both optical transitions $\ket{0}\rightarrow\ket{e}$ and $\ket{1}\rightarrow\ket{e}$,
\begin{align}\label{eq:Omega_01}
    \Omega^{0e}_q(t) &= \Omega(t) \sin(\eta_q/2), \nonumber\\
    \Omega^{1e}_q(t) &= \Omega(t) \cos(\eta_q/2) e^{i\gamma_q}.
\end{align}
The qubit dynamics can now be analyzed using the bright/dark superposition states
\begin{align}
    \ket{B_q} &= \sin(\eta_q/2) \ket{0} + \cos(\eta_q/2) e^{-i\gamma_q} \ket{1}, \nonumber \\
    \ket{D_q} &= \cos(\eta_q/2) \ket{0} - \sin(\eta_q/2) e^{-i\gamma_q} \ket{1},
\end{align}
which are respectively coupled with Rabi frequency $\Omega(t)$ and uncoupled to the excited state. The operation is therefore equivalent to that described in Sec. \ref{sec:sechyp_protocol}, except $\ket{B_q}$ and $\ket{D_q}$ assume the roles of $\ket{0}$ and $\ket{1}$, respectively. Thus, a phase $\theta$ is applied to $\ket{D_1 D_2... D_n}$, which can be set to any multi-qubit product state using the set of parameters $\{\eta_q\}$ and $\{\gamma_q\}$. 

For example, $n$-bit Toffoli gates can be implemented by using $\eta_q = \pi$ for all control qubits, $\eta_t = \pi/2$ and $\gamma_t = \pi$ for the target qubit, and $\theta = \pi$. Controlled phase gates, C$^{n-1}-$P$(\theta)$, use $\eta_q = \pi$ for all qubits, and by picking $\theta = \pi$, $\pi/2$, and $\pi/4$, one can perform C$^{n-1}-$Z, C$^{n-1}-$S, and C$^{n-1}-$T gates, respectively. One can also perform controlled rotations on the form of C$^{n-1}-e^{i\theta/2}R_{\hat{r}}(\theta)$, where $R_{\hat{r}}(\theta)$ rotates an angle $\theta$ around the vector $\hat{r}$, for more information see Appendix \ref{app:arb_gate}.

\subsection{Incorporation of single-qubit gates}
\begin{figure}
    \centering
    \includegraphics[width=\columnwidth]{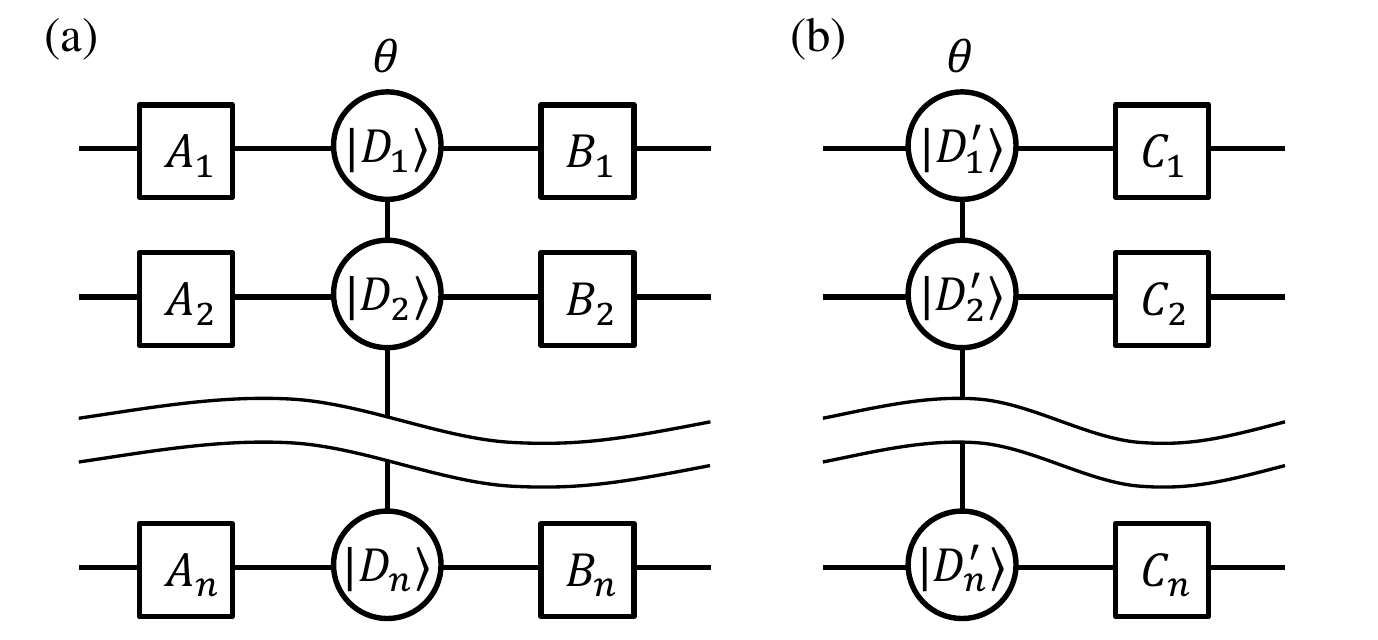}
    \caption{(a) A circuit of $n$ qubits consisting of a set of single-qubit gates $\{A_q\}$, a multi-qubit operation that adds a phase $\theta$ to $\ket{D_1 D_2 ... D_n}$, followed by another set of single-qubit gates $\{B_q\}$. (b) An equivalent circuit consisting of a multi-qubit operation that adds a phase $\theta$ to $\ket{D_1' D_2' ... D_n'}$, where $\ket{D_q'} = A_q^{-1}\ket{D_q}$, followed by a set of single-qubit gates $\{C_q\}$, where $C_q = B_q A_q$.}
    \label{fig:remove_SQ_gates}
\end{figure}

As shown in Fig. \ref{fig:remove_SQ_gates}, one can modify the multi-qubit operation to incorporate all single-qubit gates, $\{A_q\}$, that come immediately before it by changing the dark states of the operation to $\ket{D_q'} = A_q^{-1}\ket{D_q}$ and modify the gates that come afterward. Thus, if a circuit consists of purely single-qubit gates and multi-qubit gates which can be directly implemented by our gate, one can move all single-qubit gates to the end of the circuit (or, alternatively, to the beginning). Thus, if qubit measurements (or qubit initializations) can be performed on arbitrary superposition states, all single-qubit gates can be removed from the circuit. 

Assuming equal error rates for all multi-qubit operations this would reduce the total error of running the circuit. Furthermore, the removal of single-qubit gates could significantly reduce the time taken to run the algorithm, especially for rare-earth quantum computers where gates must be performed sequentially due to the dipole-dipole interactions that would otherwise occur between the qubits \cite{Kinos2021a}. 

\section{\label{sec:conc}Conclusion}
In conclusion, we have presented a robust scheme that permits a range of multi-qubit gate operations on an $n$-qubit register by application of two classical laser pulses. We assume the qubits to have distinct excitation frequencies and the pulse to have a frequency comb content with teeth at the frequency of all qubits that participate in the gate. Such frequency modulated pulses can thus be obtained from a single pulse by use of acousto- or electro-optical modulators. By forming pairs of pulses this way, so that each qubit can be excited with specific amplitude and phase from the two qubit states, the phase evolution can be made conditional on arbitrary product states of the control qubits.

We recall that the blockade gate mechanism applied for the rare-earth ions is similar to the Rydberg blockade gates mechanism applicable to neutral atoms. The latter assumes identical excitation frequencies but spatial addressing of the individual atoms in a tweezer trap array. In that system, a multi-qubit gate was proposed that uses fixed frequency pulses, and for which it was necessary to either address the atoms sequentially or assume combinations of interacting and non-interacting excited Rydberg states \cite{Isenhower2011}. Furthermore, the Rydberg atom scheme offers also other possibilities associated with the exchange of excitation among different atoms mediated by resonant dipole-dipole interactions, and single pulse adiabatic schemes for multi-qubit Toffoli and Fan-out gates have been proposed \cite{Khazali2020}, that may not find equivalent use with rare-earth ions. Whether these gates or a variant of the ones discussed in the present work will work better for neutral atoms may be pursued with the recent progress with that system \cite{Ebadi2021, Bluvstein2022, Chertkov2022, Chen2022, Graham2022}.

\begin{acknowledgments}
The authors thank Mogens Dalgaard for fruitful discussions and valuable feedback on the manuscript. This research was supported by the Swedish Research Council (Grant No. 2019-04949), the Danish National Research Foundation (Grant No. DNRF156), and it has received funding from the European Union's Horizon 2020 research and innovation program under Grants No. 820391 (SQUARE) and No. 754513 (Marie Sklodowska-Curie program).
\end{acknowledgments}

\appendix

\section{Error estimation}\label{app:error_est}
We define the error of the multi-qubit operation as 
\begin{align}\label{eq:error}
    \epsilon = 1 - \bra{\Psi_t}\rho_f \ket{\Psi_t},
\end{align}
where $\ket{\Psi_t}$ is the desired target state, which is the same as the initial state except that the state component $\ket{11...1}$ has acquired a phase of $\theta$, and $\rho_f = \ket{\Psi_f}\bra{\Psi_f}$ is the final density matrix obtained after applying the operation. 

For now, we still assume that all qubits interact with the same blockade shift given by $\Delta\omega$ and that the initial state is an even superposition of all $2^n$ computational ground states, but at the end of this section we provide an estimate for any initial state. Under these assumptions, all multi-qubit states with $n_0$ $\ket{0}$ components are equivalent, there exist $\binom{n}{n_0}$ such states, and we use $\ket{\Psi(n_0)}$ as a shorthand to indicate one of these states. 

When only considering transfer and AC Stark shift errors, the final state vector is, up to a global phase factor, estimated by 
\begin{align}\label{eq:Psi_f}
    \ket{\Psi_f} &= \frac{1}{\sqrt{2^n}}\left(\ket{11...1}e^{i\theta} + \sum_{n_0=1}^{n}\binom{n}{n_0}A(n_0)\ket{\Psi(n_0)} \right),
\end{align}
where $A(n_0)$ is defined in Eq. (\ref{eq:A_n0}) and described further in Secs. \ref{sec:transfer} and \ref{sec:AC}. 

To estimate dephasing errors the full density matrix, $\rho_f$, must be described. $\ket{11...1}$ is never excited and therefore dephasing does not impact the $\ket{11...1}\bra{11...1}$ component of $\rho_f$. For the other components we model dephasing by assuming that the first sechyp pulse excites $\ket{\Psi(n_0)}$ to a bright superposition state $\ket{B_e(n_0)}$ as described by Eq. (\ref{eq:Bright}). Following this, dephasing occurs as described in Sec. \ref{sec:dephasing}, which introduces the factors $1$, $e^{-\gamma}$, or $e^{-2\gamma}$ to the different excited density matrix components. $\gamma = \alpha t_g / T_2$ and $\alpha\approx1$ is used to estimate how large fraction of the time the qubit spends in the excited state. Lastly, the second sechyp pulse deexcites the bright superposition back to the computational state again. 

Following this procedure, the dyadic product $\ket{11...1}\bra{\Psi(n_0)}$ is excited to $\ket{11...1}\bra{B_e(n_0)}$, and since all those components dephase with the same rate of $T_2$ they all obtain the factor $e^{-\gamma}$, thus resulting in $e^{-\gamma}\ket{11...1}\bra{\Psi(n_0)}$ after the deexcitation of the second sechyp pulse.

$\ket{\Psi(m_0)}\bra{\Psi(n_0)}$, however, is excited to $\ket{B_e(m_0)}\bra{B_e(n_0)}$ which contains $n_0 m_0$ terms, where each term is on the form $\ket{e110..10}\bra{101e...11}$, i.e., the bra and ket contain exactly one excited state component. Since we assume independent dephasing of the excited states, if the $\ket{e}$ components are on different qubits a factor of $e^{-2\gamma}$ is introduced, whereas the factor is $1$ if the $\ket{e}$ components are on the same qubit. After the second sechyp pulse the dyadic product returns to \begin{align}
    \frac{k+(n_0 m_0 - k)e^{-2\gamma}}{n_0 m_0}\ket{\Psi(m_0)}\bra{\Psi(n_0)}, 
\end{align}
where $k$ is the number of qubits that start in state $\ket{0}$ in both $\ket{\Psi(n_0)}$ and $\ket{\Psi(m_0)}$, e.g., for states $\ket{011001}$ with $n_0 = 3$ and $\ket{101000}$ with $m_0 = 4$, $k = 2$, since only qubits $4$ and $5$ start in $\ket{0}$ in both states. 

The total error including all three error sources can now be estimated using Eq. (\ref{eq:error}), and the results are presented in Eq. (\ref{eq:total_error}), where the $1$ within the parenthesis comes from $\ket{11...1}\bra{11...1}$, and the first sum comes from the terms $\ket{11...1}\bra{\Psi(n_0)}$ and $\ket{\Psi(n_0)}\bra{11...1}$ of which there are $\binom{n}{n_0}$ for each $n_0 = 1,...,n$. The last term goes through all combinations of $n_0$ and $m_0$ for the states $\ket{\Psi(m_0)}\bra{\Psi(n_0)}$ and counts how many terms there exist that have $k$ qubits that start in $\ket{0}$ in both states. 

In order to estimate the error for any initial state, we simplify the calculations by making the assumption that $\ket{e0}$ dephase with $\ket{e1}$ with a rate of $T_2/2$. This increases the estimated error due to dephasing if $n$ is low, but has little impact if $n$ is large. The error then becomes
\begin{widetext}
\begin{align}\label{eq:total_error_all_init}
    \epsilon &= 1 - \bigg(P(0)^2 + P(0)\sum_{n_0=1}^{n} P(n_0) 2 e^{-\gamma} \text{Re}\left[A(n_0)\right] + \sum_{n_0=1}^{n}\sum_{m_0=1}^{n} P(n_0) P(m_0) \text{Re}\left[A(n_0)A^*(m_0)\right] e^{-2\gamma}\bigg), \nonumber\\
    P(n_0) &= \sum_{i=1}^{\binom{n}{n_0}}|a_i(n_0)|^2, 
\end{align}
\end{widetext}
where $P(n_0)$ is the probability to be in any of the states with $n_0$ $\ket{0}$ components, and $a_i(n_0)$ is the complex amplitude of being in one of the $\binom{n}{n_0}$ different $\ket{\Psi(n_0)}$ states. On average, $P(n_0) = \binom{n}{n_0} \frac{1}{2^n}$ and in that case the error is estimated by Eq. (\ref{eq:total_error}).

\section{\label{app:generalization_AC_err}Estimating the AC Stark shift error}
In the absence of dephasing, the evolution of an initial state $\ket{\Psi(n_0)}$ can be obtained by solving the Schr\"odinger equation for the complex state amplitudes, 
\begin{align}\label{eq:state_amplitudes}
    c_{\Psi} &= -\frac{i\Omega}{2} \left(\sum_{q=1}^{n_0} c_{q}\right), \nonumber \\
    c_{q} &= -\frac{i\Omega}{2} \left(c_{\Psi} + \sum_{p=1, p\neq q}^{n_0} c_{qp}\right), \nonumber \\
    c_{qp} &= -\frac{i\Omega}{2} \left(c_{q} + c_{p}\right) - i \Delta\omega_{qp} c_{qp},
\end{align}
where $c_{q}$ ($c_{qp}$) is the complex state amplitude of the state where qubit $q$ (qubits $q$ and $p$) is excited and all other qubits are in their initial state, and $\Delta\omega_{qp}$ is the blockade shift between qubits $q$ and $p$. Both $q$ and $p$ are used to iterate through the $n_0$ qubits that start in $\ket{0}$. To simplify the notation we do not write out any time dependence and we have assumed that $\Omega$ is real. Furthermore, we have assumed that we can neglect the drive to any triply excited state. 

We will now rewrite this system of equations using symmetrized state amplitudes
\begin{align}\label{eq:Be_Bee_states}
    c_{Be} &= \frac{1}{\sqrt{n_0}} \sum_{q=1}^{n_0} c_{q}, \nonumber \\
    c_{Bee} &= \frac{1}{\sqrt{n_{ee}}} \sum_{q=1}^{n_0}\sum_{p=q+1}^{n_0} c_{qp},
\end{align}
where $n_{ee} = \binom{n_0}{2} = \frac{n_0(n_0-1)}{2}$. If all shifts are equal, $\Delta\omega_{qp} = \Delta\omega$, we obtain the system shown in Fig. \ref{fig:sechyp_protocol}(b). 

\begin{figure*}
    \centering
    \includegraphics[width=\textwidth]{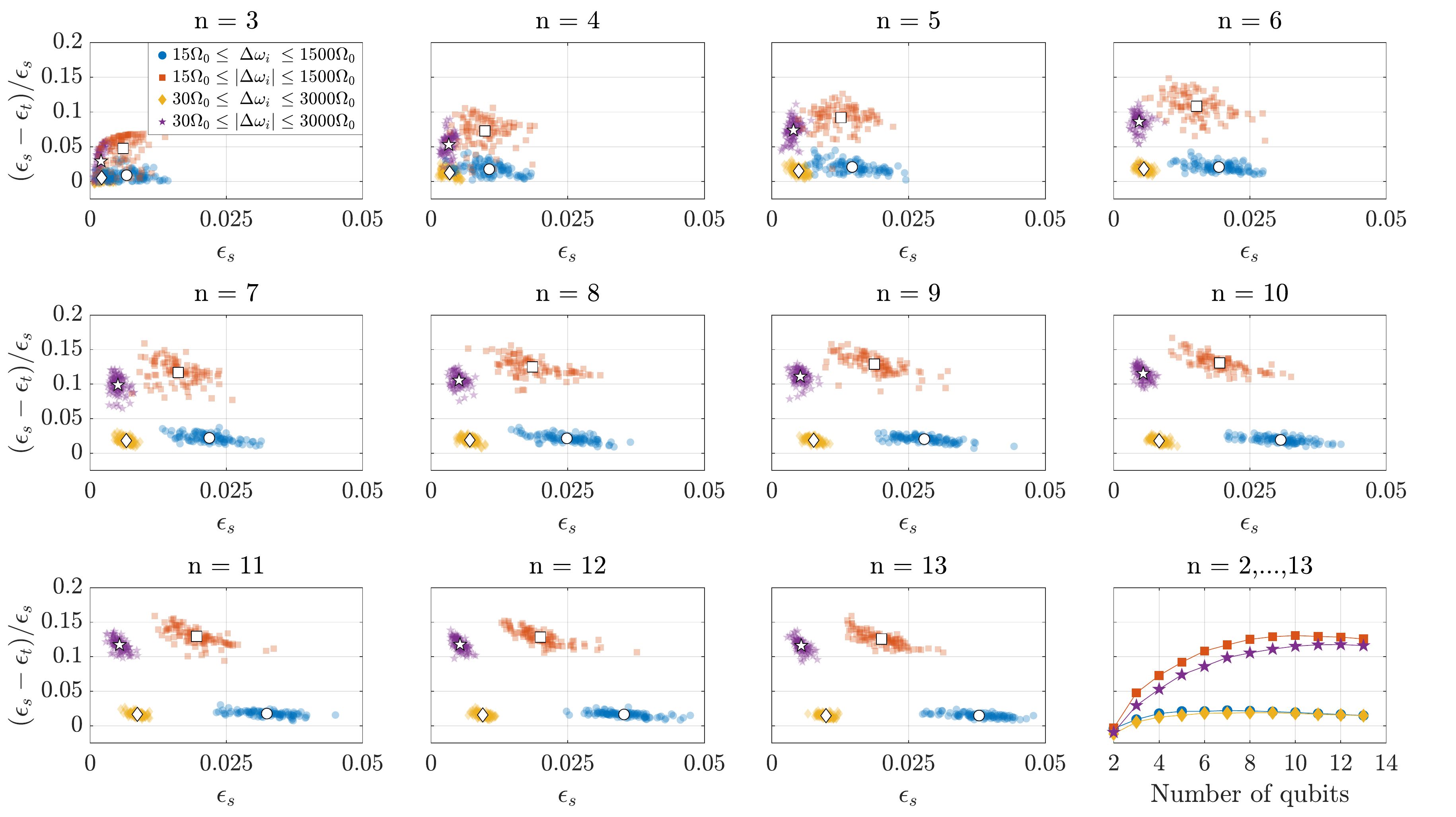}
    \caption{This figure shows the relative deviation between the theoretically estimated error $\epsilon_t$ and the numerically simulated error $\epsilon_s$ as a function of $\epsilon_s$ for $n=3,...,13$ qubits. For each $n$ we run $100$ different simulations with random shifts $\Delta\omega_{qp}$ drawn from the distributions described in Sec. \ref{sec:arb_shifts}. The blue circles and yellow diamonds (red squares and purple stars) show the results when the shifts are only positive (positive and negative), and the white markers show the mean relative deviation at the mean simulated error. The last panel shows the mean relative deviation as a function of the number of qubits for $n = 2,..., 13$.}
    \label{fig:theory_deviation}
\end{figure*}

However, in the general case where $\Delta\omega_{qp}$ are different for each pair of qubits, $\ket{B_{ee}}$ is also coupled to other doubly excited states. We therefore define a full set of doubly excited superpositions states $\{\ket{c^{(k)}}\}$, $k = 1,...,n_{ee}$ where $\ket{c^{(1)}} = \ket{B_{ee}}$. To be able to estimate the AC Stark shift error we neglect any drive between singly and doubly excited states except for the drive between $\ket{B_e}$ and $\ket{B_{ee}}$, which is acceptable in most cases since their strength is proportional to $\Omega_0$, whereas the interactions between different doubly excited states scale with $\Delta\omega_{qp} \gg \Omega_0$. Furthermore, we define our basis set such that $\ket{c^{(k)}}$ only couples to state $\ket{c^{(k-1)}}$ and $\ket{c^{(k+1)}}$. Thus, Eq. (\ref{eq:state_amplitudes}) can be written as
\begin{align}
    c_{\Psi} &= -\frac{i\sqrt{n_0}\Omega}{2} c_{Be}, \nonumber \\
    c_{Be} &= -\frac{i\sqrt{n_0}\Omega}{2} c_{\Psi} -\frac{i\sqrt{2(n_0-1)}\Omega}{2} c^{(1)}, \nonumber \\
    c^{(1)} &= -\frac{i\sqrt{2(n_0-1)}\Omega}{2} c_{Be} - \frac{i\Omega^{(2)}}{2} c^{(2)} - i \Delta\omega^{(1)} c^{(1)},  \nonumber \\
    c^{(k)} &= - \frac{i\Omega^{(k)}}{2} c^{(k-1)} - \frac{i\Omega^{(k+1)}}{2} c^{(k+1)} - i \Delta\omega^{(k)} c^{(k)},  \nonumber \\
    k &= 2,..., n_{ee}-1, \nonumber\\
    c^{(n_{ee})} &= - \frac{i\Omega^{(n_{ee})}}{2} c^{(n_{ee}-1)} - i \Delta\omega^{(n_{ee})} c^{(n_{ee})}, 
\end{align}
where
\begin{align}\label{eq:bright_dark}
    \Omega^{(k)} &= 2\sum_{q=1}^{n_0}\sum_{p=q+1}^{n_0} d_{qp}^{(k-1)}d_{qp}^{(k)}\Delta\omega_{qp}, \nonumber \\
    \Delta\omega^{(k)} &= \sum_{q=1}^{n_0}\sum_{p=q+1}^{n_0} \left(d_{qp}^{(k)}\right)^2\Delta\omega_{qp}, \nonumber \\
    c^{(k)} &= \sum_{q=1}^{n_0}\sum_{p=q+1}^{n_0} d_{qp}^{(k)}c_{qp}. 
\end{align}
The $d_{qp}^{(k)}$ coefficients relate $\ket{c^{(k)}}$ to $\ket{c_{qp}}$, and for $\ket{c^{(1)}} = \ket{B_{ee}}$ the coefficients are given by Eq. (\ref{eq:Be_Bee_states}), i.e., $d_{qp}^{(1)} = 1/\sqrt{n_{ee}}$. For the other states with $k = 2,...,n_{ee}$ the coefficients are calculated using 
\begin{align}
    d_{qp}^{(k)} &= \frac{1}{\sqrt{D}} \left( f_{qp}^{(k)} - \sum_{s=1}^{k-1} d_{qp}^{(s)} \left(\sum_{q=1}^{n_0}\sum_{p=q+1}^{n_0} d_{qp}^{(s)} f_{qp}^{(k)} \right)\right),  \nonumber \\
    f_{qp}^{(k)} &= \frac{1}{\sqrt{F}} d_{qp}^{(k-1)} \Delta\omega_{qp},
\end{align}
where $D$ and $F$ are normalization factors calculated through 
\begin{align}
    \sum_{q=1}^{n_0}\sum_{p=q+1}^{n_0} \left(d_{qp}^{(k)}\right)^2 &= 1, \nonumber \\
    \sum_{q=1}^{n_0}\sum_{p=q+1}^{n_0} \left(f_{qp}^{(k)}\right)^2 &= 1.
\end{align}

To estimate the AC Stark shift error we still use Eqs. (\ref{eq:total_error}) and (\ref{eq:A_n0}), except $\Delta\omega$ now depend on the initial state $\ket{\Psi(n_0)}$ and is replaced by an effective shift $\Delta\omega_{\text{eff}}^{(1)}$ of $\ket{B_{ee}}$ which is calculated through the recursive equation
\begin{align}\label{eq:recursive}
    \Delta\omega_{\text{eff}}^{(k)} &= \Delta\omega^{(k)} - \frac{\left(\Omega^{(k+1)}\right)^2}{4\Delta\omega^{(k+1)}_{\text{eff}}}, \nonumber\\
    \Delta\omega_{\text{eff}}^{(n_{ee})} &= \Delta\omega^{(n_{ee})}. 
\end{align}

In Fig. \ref{fig:theory_deviation} we verify that our estimated error agrees with the numerically simulated error, since even though the deviation initially grows as a function of $n$ it eventually saturates.

\section{\label{app:simulation}Simulations}
The simulations were performed by evolving the Lindblad master equation (or the Schr\"odinger equation when dephasing was not included) using MATLAB's explicit Runge-Kutta ode45 function \cite{Dormand1980, Shampine1997}, where the relative and absolute tolerances were set to $10^{-10}$ ($10^{-8}$ for the simulations presented in Fig. \ref{fig:random_shifts_err}). Multi-qubit states containing three or more excited state components (two or more when running without AC Stark shift errors) were not included in any simulations, but the errors due to driving these levels are small.

For the case without dephasing and equal blockade shifts in Fig. \ref{fig:sechyp_errors}(a), the symmetric bright states of Eq. (\ref{eq:Be_Bee_states}) are simulated together with two ground states $\ket{0}$ and $\ket{1}$ for each subsystem $\ket{\Psi(n_0)}$ with $n_0 = 1,...,n$, since all $\binom{n}{n_0}$ different $\ket{\Psi(n_0)}$ states are equivalent. Since only four levels are required for $n$ different simulations, results can be obtained for all $n=2,...,50$. 

Dephasing was modeled using the Lindblad master equation 
\begin{align} \label{eq:LM}
\frac{d\rho}{dt}&=\frac{1}{i\hbar}[H,\rho] - \nonumber\\
    & \frac{1}{2}\sum_{m=1}^n (C_m^\dagger C_m \rho + \rho C_m^\dagger C_m) + \sum_{m=1}^n C_m \rho C_m^\dagger,
\end{align}
with operators 
\begin{align}
    C_m = \left(\prod_{i=1}^{m-1} I\right) \otimes C \otimes \left(\prod_{j=m+1}^{n} I\right), \nonumber \\
    C = \frac{1}{\sqrt{2T_2}} (\ket{e}\bra{e} - \ket{0}\bra{0} - \ket{1}\bra{1}),
\end{align}
where $I$ is the identity operator and $T_2$ is the coherence time of the excited state. Since this system grows exponentially with the number of qubits, we have only simulated the cases where $n=2,...,8$ in Fig. \ref{fig:sechyp_errors}(b-c). 

The transfer error factor $T(n_0)$ in Eq. (\ref{eq:A_n0}) was calculated by evolving the Schr\"odinger equation for a two-level system, $\ket{g}$ and $\ket{e}$, with initial state $\ket{g}$ when two consecutive sechyp pulses first excited and then deexcited the system. The second sechyp pulse used $\theta = \pi$, i.e., it had the same phase as the first pulse since $\pi + \theta = 2\pi$. $T(n_0)$ was then set equal to the final complex amplitude of the ground state multiplied by $e^{i\theta}$. The dependence on $n_0$ was implemented by repeating the simulation for different Rabi frequencies scaled by $\sqrt{n_0}$, $n_0 = 1,...,n$. 

When optimizing $\Omega_0$ and the cutoff duration in Fig. \ref{fig:sechyp_errors}(c) to minimize the error of Eq. (\ref{eq:total_error}), MATLAB's fminsearch function was used. In this case, $T(n_0)$ also depends on the cutoff duration used, and 5000 simulations were performed using Rabi frequency factors of $\sqrt{n_0}$, $n_0 = 1,...,50$, and 100 equally sampled points between $t_g/t_\text{fwhm} = 2,...,10$. $T(n_0, t_g/t_\text{fwhm})$ was then estimated for any factor $t_g/t_\text{fwhm}$ by using MATLAB's interp2 function with bilinear interpolation.

\section{\label{app:arb_gate}Multi-qubit controlled rotations}
An arbitrary single-qubit gate operation, $U$, can be written as \cite{Nielsen2010}:
\begin{align}\label{eq:SQ_op}
    &U = e^{i\alpha}R_{\hat{r}}(\theta) = e^{i\alpha}e^{-i\theta \hat{r}\cdot \overrightarrow{\sigma}/2}, \\
    &= e^{i\alpha}\big(\cos(\theta/2)I - i \sin(\theta/2) \left(r_x X + r_y Y + r_z Z\right)\big), \nonumber
\end{align}
where $R_{\hat{r}}(\theta)$ denotes a rotation around vector $\hat{r} = (r_x, r_y, r_z)$ with an angle $\theta$, $\alpha$ is a global phase, and $\overrightarrow{\sigma}$ is the three component vector $(X, Y, Z)$ of the Pauli matrices in the computational $\{\ket{0}, \ket{1}\}$ basis. 

The multi-qubit sechyp operation adds a phase $\theta$ to the $\ket{D_1 D_2...D_n}$ state, which we can analyze from the point of view of assigning qubits $1,...,n-1$ as controls and qubit $n$ as the target. If all controls are in their respective dark states the gate performs the following operation on the target 
\begin{align}
    &U' = e^{i\theta}\ket{D_n}\bra{D_n} + \ket{B_n}\bra{B_n}, \\ 
    &= e^{i\theta/2} \left(\cos(\theta/2) I - i\cdot \sin(\theta/2) \left(r_x X + r_y Y + r_z Z \right)\right), \nonumber
\end{align}
where
\begin{align}
    r_x &= 2\sin{\left(\eta_n/2\right)}\cos{\left(\eta_n/2\right)} \cos(\gamma_n), \nonumber \\
    r_y &= -2\sin{\left(\eta_n/2\right)}\cos{\left(\eta_n/2\right)} \sin(\gamma_n), \nonumber \\
    r_z &= 2\sin{\left(\eta_n/2\right)}^2 - 1.
\end{align}
Comparing this to Eq. (\ref{eq:SQ_op}) the only difference is in the phase factors $e^{i\alpha}$ and $e^{i\theta/2}$, since $\eta_n$ and $\gamma_n$ can be used to set any direction $\hat{r}$ and we can apply any $\theta$. Thus, an additional phase of $\alpha' = \alpha - \theta/2$ should be applied to the state $\ket{D_1 D_2...D_{n-1}} \otimes I$ in order to perform an arbitrary gate on the target. If $\alpha'$ is a multiple of $2\pi$, the operation can be implemented directly using only one multi-qubit sechyp operation. The general case can be done using two multi-qubit operations: first, the operation analyzed above, and second, an operation on only the $n-1$ control qubits to apply a phase $\alpha'$ to $\ket{D_1 D_2...D_{n-1}} \bra{D_1 D_2...D_{n-1}}\otimes I_n$. Alternatively, one could modify the multi-qubit operation: first, apply one sechyp pulse on all controls; second, perform an arbitrary single-qubit gate on the target (see, e.g., \cite{Kinos2021a, Kinos2022b}) which is only applied if all controls are in their respective dark states, and, similarly to the analysis above, lacks a phase $\alpha'$; third, deexcite all controls using a phase of $\pi+\alpha'$ which adds a phase $-\alpha'$ on all states except $\ket{D_1 D_2...D_{n-1}} \bra{D_1 D_2...D_{n-1}}\otimes I$, which up to a global phase factor is equivalent to adding a phase $\alpha'$ to $\ket{D_1 D_2...D_{n-1}}\bra{D_1 D_2...D_{n-1}} \otimes I$.

\bibliography{Ref_lib}

\end{document}